\documentstyle[12pt]{article}
\newcommand{\be}{\begin{equation}}
\newcommand{\ee}{\end{equation}}
\newcommand{\bea}{\begin{eqnarray}}
\newcommand{\eea}{\end{eqnarray}}
\begin{document}
\baselineskip 18pt
\begin{flushright}{ UTPT--98--05}
\end{flushright}
\begin{center}
{\Large \bf Scale of Leptogenesis}\\
\vskip .3in
{\bf Jacqueline Faridani$^{(a)}$, {\footnote 
{jafarida@mercator.math.uwaterloo.ca}} 
Smaragda Lola$^{(b)}$, {\footnote{magda@mail.cern.ch}}
\hbox{Patrick J. O'Donnell$^{(c)}$} {\footnote{pat@medb.physics.utoronto.ca}}
and Utpal Sarkar$^{(d)}$} {\footnote{utpal@prl.ernet.in}} \\
\vspace{2cm}
(a) Department of Statistics and Actuarial Science,
University of Waterloo,
Waterloo, Ont. N2L 3G1, Canada.\\
(b) Theory Division, CERN, CH-1211 Switzerland, Geneva\\
(c) Physics Department, University of Toronto, Toronto, Ontario M5S 1A7,
Canada  \\
(d)Theory Group, Physical Research Laboratory,
Ahmedabad, 380 009, India\\
\end{center}

\vskip  .5in  
\begin{abstract}  
\baselineskip  15pt 

We study the scale at which one can generate the lepton asymmetry
of the  universe  which  could  then get  converted  to a  baryon
asymmetry during the electroweak  phase  transition.  We consider
the  possibility   that  the  Yukawa   couplings  are  small  but
sufficiently  large to generate  enough  lepton  asymmetry.  This
forbids the  possibility of the $(B-L)$  breaking scale being the
electroweak  scale.  

\end{abstract}  

\newpage
\baselineskip  16pt 

\section{Introduction}

In most grand unified theories (GUTs) the baryon asymmetry of the
universe   is   generated   during   the  GUT  phase   transition
\cite{sak,kolb,gut}.  In these  models  the  generated  asymmetry
also implies an equal amount of lepton  asymmetry and hence there
is  no  net   $(B-L)$   asymmetry.  On  the  other  hand  if  the
electroweak phase transition is a second order phase  transition,
then any primordial  $(B+L)$  asymmetry  generated during the GUT
phase transition will be washed out \cite{krs} .

This  situation can be saved in models where $(B-L)$ is broken at
some intermediate  scales.  In this case a $(B-L)$  asymmetry can
be generated through higgs decay or heavy Majorana neutrino decay
if there is appropriate $CP$ violation \cite{fy}--\cite{paschos}.
The  out-of-equilibrium  condition  then imposes a lower bound on
this symmetry breaking scale to be around $10^7$ GeV.  This bound
is  dependent  on the fact that the Yukawa  couplings  are larger
than  $10^{-5}$.  Although   esthetically   this  number  sounds
reasonable, nothing tells us definitely that the Yukawa couplings
cannot  be  smaller   than  this.  For  example,  if  the  Yukawa
couplings   relating  the  left  handed   leptons  to  the  first
generation right handed heavy neutrinos are similar to the Yukawa
couplings for the electron, then the out-of-equilibrium condition
can be  satisfied  for even a TeV scale for  left-right  symmetry
breaking.  But the same Yukawa  couplings enter in the expression
for the  generated  $(B-L)$  asymmetry,  which  may  then be very
small.  In this  article we study  systematically  the  Boltzmann
equations for the  generation  of the lepton  asymmetry and hence
$(B-L)$  asymmetry  to find out the  lowest  possible  left-right
symmetry  breaking scale which  satisfies the  out-of-equilibrium
condition  and  generates   enough  baryon  asymmetry  after  the
electroweak phase transition.

In the next section we briefly review the leptogenesis  scenario,
where one  generates  a lepton  asymmetry  when the right  handed
Majorana  neutrinos  decay.  This then gets converted to a baryon
asymmetry during the electroweak phase transition.  Subsequently,
we discuss the Boltzmann equations and the possible solutions for
a low energy  left-right  symmetry  breaking. 

\section{Model for leptogenesis}

It was first proposed by Fukugita and Yanagida  \cite{fy} that in
extensions of the standard  model, which  include  singlet  heavy
right  handed  neutrinos,  it is  possible  to  generate a lepton
asymmetry at some intermediate  scale ($\sim 10^{10}$ GeV), which
can  then  get  converted  to  a  baryon  asymmetry   during  the
electroweak  phase  transition.  The source of CP violation is in
the  mass  matrix  of  the   neutrinos.  The   out-of-equilibrium
condition  is  satisfied  with small  Yukawa  couplings.  In this
scenario  the heavy right  handed  neutrinos  decay to light left
handed neutrinos in  out-of-equilibrium.  The amount of asymmetry
thus generated depends on the amount of CP violation.

In this  approach it is  expected  that at the time of the decay,
there  are  enough  right   handed   neutrinos   given  by  their
equilibrium  distribution.  However, in practice it is  difficult
to achieve this distribution  \cite{plum}.  Even though we assume
that gravitational  fields have generated  equilibrium numbers of
right handed  neutrinos,  during the period  $10^{18}$  GeV $\to$
$10^{10}$ GeV the equilibrium  distribution  cannot be maintained
since the right handed  neutrinos are sterile  under the standard
model and as the  universe  expands the number  density  reduces.
When   these   fewer   right    handed    neutrinos    decay   in
out-of-equilibrium, they cannot generate enough lepton asymmetry.

It has been pointed out that if these right handed neutrinos take
part in any  gauge  interaction,  then the  decays  of the  gauge
particles can produce an  equilibrium  distribution  of the right
handed  neutrinos.  The  left-right   symmetric  models  are  the
natural  extensions for this, where the heavy neutrinos  interact
with the right  handed gauge bosons  strongly  \cite{plum}.  As a
result, the decay of the right  handed gauge  bosons can generate
an equilibrium  distribution of heavy  neutrinos.  In addition to
the heavy neutrino  decays, lepton number  violating higgs decays
also  contribute to the  generation of lepton  asymmetry in these
models \cite{pat}.

In the  following we shall thus  consider a left-right  symmetric
extension  of  the  standard  model.  We  consider  the  symmetry
breaking chain,
$
SU(3)_c \times SU(2)_L \times 
SU(2)_R \times U(1)_{(B-L)}$ $ \left[ \equiv G_{LR} \right]
{M_R \atop \longrightarrow}$ $ SU(3)_c \times SU(2)_L  \times
U(1)_Y  \left[ \equiv G_{std} \right] $ $
{M_W \atop \longrightarrow}$ $ SU(3)_c \times U(1)_{em}.$
The symmetry  breaking  $G_{LR} \to G_{std}$ takes place when the
right handed  triplet higgs field  $\Delta_R  \equiv$  (1,1,3,-2)
acquires a vacuum  expectation  value ({\it vev}).  In this model
$(B-L)$ is a local  symmetry.  The breaking of the group $G_{LR}$
also implies spontaneous  breaking of $(B-L)$.  Left-right parity
implies the  existence of another  higgs field  $\Delta_L $ which
transforms  as  (1,3,1,-2)  under  $G_{LR}$.  A higgs  bi-doublet
field $\phi$ (1,2,2,0) breaks the electroweak  symmetry and gives
masses to the fermions.

The fermion content of the model is, $q_{iL} \equiv [3,2,1,1/3]$, 
$q_{iR} \equiv [3,1,2,1/3]$, $\ell_{iL} \equiv [1,2,1,1]$ and
$\ell_R \equiv [1,1,2,1]$, where $i = 1,2,3$ corresponds to three 
generations. The right handed neutrinos ($N_i \equiv \nu_{iR}$) are 
contained in $\ell_{iR}$ and we do not have to include them by hand.
The Yukawa couplings in the leptonic sector are given by,
\begin{equation}  
{\cal  L}_{Yuk} = f_{ij}  \overline{\ell_{iL}}  \ell_{jR}  \phi 
+ f_{Lij}  \overline{{\ell_{iL}}^c}  \ell_{jL}  \Delta_L^\dagger  
+ f_{Rij}  \overline{{\ell_{iR}}^c}  \ell_{jR} \Delta_R^\dagger.
\label{Yuk}
\end{equation} 

The scalar potential has many more terms compared to the standard
model.  We write down only those  terms which  contribute  to the
generation of the lepton asymmetry of the universe,
\begin{equation}  
{\cal L}_{int} = g  (\Delta_L^\dagger  \Delta_R \phi \phi 
+ \Delta_L  \Delta_R^\dagger \phi \phi ) + h.c. . \label{scalar}
\end{equation}  
The  {\it  vev}\,s  of  these  fields  are  not  independent.  We
consider the minimum of the complete  potential  which  satisfies
$v_L \ll v_R$ and $v_L \approx {v^2 / v_R}$, where  $v_{L,R}$ and
$v$ are the {\it vev}\,s of the fields  $\Delta_{L,R}$ and $\phi$
respectively.  We also assume that the  left-right  parity  ($D-$
parity)  is not  broken,  and  hence  the  masses  of the  fields
$\Delta_L  $ and  $\Delta_R$  remains  the same  even  after  the
breaking of $G_{LR}$, {\it i.e.,}  $m_{\Delta_L} = m_{\Delta_R} =
m_{\Delta}  \approx  v_R$.  At this  scale  $v_R$,  the $(B - L)$
local  symmetry is also broken by two units,  which gives rise to
Majorana   masses  of  the   neutrinos  and   neutron-antineutron
oscillations.

The   $\Delta_{L,R}$  can  now  decay  into  two  leptons,  while
$\Delta_{L,R}^\dagger$ decay into two antileptons:
\begin{eqnarray}
  \Delta_{L,R}  &\to& \ell_{L,R} + \ell_{L,R}. \label{eqna} \\
 \Delta_{L,R}^\dagger  &\to& \ell_{L,R}^c + \ell_{L,R}^c. \label{eqnb}
\end{eqnarray} 
These interactions, along with the scalar interactions 
$$ \Delta_{L,R} \to \phi + \phi,
$$ $$ \phantom{\Delta_{L,R}}  \to \phi^\dagger + \phi^\dagger,$$
give rise to lepton number  violation.  The  interference  of the
tree level  diagram and the one loop  diagram of Fig.  1 can then
give  rise  to  a  lepton   asymmetry   in  the  decay  modes  of
(\ref{eqna})  and  (\ref{eqnb})  for  the  left-handed   triplets
$\Delta_L$ given by,
\begin{equation}
\epsilon_\Delta  \approx  \frac{1}{8  \pi  |f_{Lij}|^2}  
{\rm  Im}  [g^* f_{Lij}^*  f_{ik}  f_{jk}]  
F \left(\frac{g^* }{f_{Rkk}}\right) , \label{eps}
\end{equation}
where  $F(q)  =  {\rm  ln}  (1  +  1/q^2)$.  The  quantity  $[g^*
f_{Lij}^* f_{ik} f_{jk}]$  contains a $CP$ violating phase and so
can be complex.

Although the left handed higgs triplets  $\Delta_L$  can generate
lepton   asymmetry   through   their  lepton   number   violating
interactions, the scattering  process  \hbox{$\Delta_L + \Delta_L
\to W_L + W_L$} will make their  number  density  the same as the
equilibrium density below the GUT scale so, at low energies, they
cannot  satisfy  the  out-of-equilibrium  condition.  So for  the
generation   of  lepton   asymmetry   at  low   energy  the  main
contribution  comes from the heavy right handed neutrino  decays.
The $vev$ of  $\Delta_R$  spontaneously  gives a Majorana mass to
the right  handed  neutrinos.  This in turn  allows  the decay of
$\nu_R$ into a lepton and an antilepton,
\begin{eqnarray}
  N_i  &\to& \ell_{jL} + \bar{\phi}, \label{eqn1a} \\
   &\to&  {\ell_{jL}}^c + {\phi} .\label{eqn1b}
\end{eqnarray}
 In the case of decays of the right-handed neutrinos there are two
types of loop  diagrams  which can interfere  with the tree level
decays of (\ref{eqn1a}) and (\ref{eqn1b}) which are shown in Fig.
2.  The  interference  of the tree level diagram and the one loop
diagrams of Fig.  2 generates a lepton asymmetry given by,
\begin{equation}
\epsilon_\nu  \approx  \frac{1}{4  \pi  |f_{ik}|^2}  {\rm  Im}  [f_{ik}
f_{il}  f_{jk}^*  f_{jl}^*]  \frac{f_{Rii}}{f_{Rkk}} .\label{eps1}
\end{equation}
In addition to these one loop vertex type corrections,  there are
self  energy  type  corrections   \cite{ls,paschos}   which  also
contribute  to the lepton  asymmetry of the  universe  (Fig.  3).
This  contribution  can also be  viewed  as a new  indirect  $CP$
violation  entering  in the mass  matrix  of the heavy  neutrinos
\cite{paschos}.  This contribution becomes much stronger than the
other   contribution   when  the  heavy   neutrinos   are  almost
degenerate.

When the mass squared  difference  between the two generations of
heavy neutrinos are large, the interference of the tree level and
the one  loop  diagrams  of Fig.  3 has  already been  calculated  in the
literature \cite{ls}.  This contribution has also been calculated
in the other  approach,  where  this is viewed as a new  indirect
$CP-$violation  $\delta$,  which enters  through the mass matrix.
Both the results  come out to be the same and that  confirms  the
equivalence  between  the two  approaches.  However,  the  latter
approach has the  advantage  that it allows us to  calculate  the
amount  of $CP$  violation  in the  limit  when the mass  squared
difference is very small.

In the case of a small mass difference, $\delta$ reads 
\cite{paschos}:
\begin{equation}
\delta  =   2 \, \pi \, g^{ab} {\cal C} \frac{M_1 M_2}{M_2^2 - M_1^2} 
\end{equation}
where
\begin{equation}
{\cal C} = - {1 \over \pi}
  {\rm Im}[ \sum_\alpha (f_{\alpha 1}^\ast f_{\alpha 2})
 \sum_\beta f_{\beta 1}^\ast f_{\beta 2})] \left( 
         \frac{1}{\sum_{\alpha} |f_{\alpha 1}|^2}
         + \frac{1}{\sum_{\alpha} |f_{\alpha 2}|^2 } \right)
\end{equation}
This   contribution   becomes   significant  when  the  two  mass
eigenvalues  are close to each other.  It  indicates  a resonance
like behaviour of the asymmetry if the two mass  eigenvalues  are
nearly  degenerate.  For very large values of the mass difference
the two contributions $\epsilon_\nu$ and $\delta$ are of the same
order of magnitude.

We now have to consider only the  $\Delta_L,  \; \Delta_R \; {\rm
and} \;  N_{1,2}$  decay  processes.  We  assume  $N_1$ to be the
lightest of the right  handed  neutrinos.  If the masses of $N_1$
and $N_2$ are almost  degenerate,  their decay  widths can become
larger than the mass  difference.  In this case both the neutrino
decays will  contribute to the lepton  asymmetry of the universe.
The decay widths for $\Delta_{L,R}$ and $N_1$ are,
\begin{equation}
\Gamma_{\Delta_{L,R}} =  \displaystyle \frac{| f_{[L,R]ij} |^2}{16 \pi} 
 M_\Delta \hskip .4in {\rm and  } \hskip .4in
\Gamma_{N_i} =  \displaystyle \frac{| f_{1j} |^2}{16 \pi} 
 M_N,  
\end{equation}
where  $M_N$  is the  mass of  $N_1$.  Since  we  assumed  $M_N <
M_\Delta$,  at low energy the  $\Delta_L$  decay  will  erase all
lepton  asymmetry  and then the $N_1$  decay  will  generate  the
required asymmetry.  For this reason while working the details of
the  Boltzmann  equation  we take the effect of only  $\epsilon =
\epsilon_\nu$  and  $\delta$.  We shall now  proceed to solve the
Boltzmann  equations  including all these  contributions  and the
scattering processes.

\section{Solutions of the Boltzmann equations}

The evolution of lepton and neutrino densities is governed by the
Boltzmann equations.  We start by deriving the Boltzmann equation
for the neutrino  number density  $n_i$.  The equation  governing
the evolution of $n_i$ is:\cite{kolb}
\bea
\dot{n_i}+3Hn_i&=&\int\!\!{\rm d}\Pi_i{\rm d}\Pi_1{\rm d}\Pi_2\;(2\pi)^4
\delta^{(4)}(p_i-p_1-p_2)\nonumber\\
&&\times \left\{ -f_i(p_i)|{\cal M}_0|^2+{1\over 2}(1+\epsilon)
|{\cal M}_0|^2 f_l(p_1)f_{\Phi^c}(p_2)\right. \nonumber\\
&&\left. +{1\over 2}(1-\epsilon)|{\cal M}_0|^2 f_{\bar l}(p_1)f_\Phi(p_2)
\right\} \nonumber\\
&=&\int\!\!{\rm d}\Pi_i{\rm d}\Pi_1{\rm d}\Pi_2\;(2\pi)^4
\delta^{(4)}(p_i-p_1-p_2)\nonumber\\
&&\left \{ -f_i(p_i)+f_i^{eq}(p_i)\right \}|{\cal M}_0|^2+{\cal
O}(\epsilon,\mu/T)\nonumber\\
&=&-\Gamma_i(n_i-n^{eq}_i),
\eea
where $n_i^{eq}$ is the equilibrium  number density of the $N_i$,
and  $\Gamma_i$ is the thermally  averaged  decay width of $N_i$.
The term on the left-hand  side is the time  derivative of $n_i$,
plus a term  which  accounts  for  the  dilution  effect  of  the
expansion of the universe.  The  integration  is over phase space
${\rm d}\Pi$ and the phase space  densities $f$, are given by the
Maxwell-Boltzmann statistics:
\be
f_i(E)=\exp [-{(E-\mu_i)\over T}],\;\;f_l(E)=\exp[-{(E-\mu)\over T}],\;\; 
f_\Phi(E)=\exp [-{(E+\mu)\over T}],
\ee
where $\mu$ is the chemical potential. 
The matrix element ${\cal M}_0$ is defined by:
\bea
|{\cal M}(N\to {\bar l}\Phi)|^2&=&|{\cal M}(l\Phi^c\to N)|^2=
{1\over 2}(1+\epsilon)|{\cal M}_0 |^2,\nonumber\\
|{\cal M}(N\to l\Phi^c)|^2&=&|{\cal M}({\bar l}\Phi\to N)|^2
={1\over 2}(1-\epsilon)|{\cal M}_0|^2,
\eea
where $\epsilon$ is a measure of $CP$-violation.

It is more convenient to work with the variables:
\bea
Y_i=n_i/s, && x=M_i/T=[2H(x=1)t]^{1\over 2},
\eea
where  $M_i$  is the  mass of  $N_i$; $s = g_* n_\gamma$ is the entropy
density of the universe; $g_*$ is the total spin degrees of freedom; 
$n_\gamma$ is the equilibrium photon density
of the universe  and  $Y_i$  the  number  of
neutrinos per co-moving volume element.

Thus: 
\be
{{\rm d}Y_i\over{\rm d}x}=-K\gamma x(Y_i-Y_i^{eq}),
\ee
where we have used:
\bea
n_\gamma=s/g_*, && {{\rm d}s\over{\rm d}t}=-3s{{\dot R}\over R}=-3sH,
\eea
and:
\bea
K={\Gamma_i(x=1)\over H(x=1)}, && \gamma={\Gamma_i(x)\over\Gamma_i(x=1)},
\eea
with: 
$$
Y_i^{eq}=n_i^{eq}/s=\left \{
\begin{array}{ll}
g_*^{-1} & x\ll 1\\
g_*^{-1}\sqrt{\pi/2}x^{3/2}\exp(-x) & x\gg 1
\end{array}
\right. .
$$
In solving the Boltzmann equations, we make the further change of
variables, $X=g_*Y$, thus:
\be
{{\rm d}X_i\over{\rm d}x}=-K\gamma x(X_i-X_i^{eq}).
\ee
This is the  Boltzmann  equation  for the  evolution  of neutrino
number density, with the initial condition $X_i(0)=1$.

We now derive the Boltzmann  equation for  $L={1\over  2}(l-{\bar
l})$,  where  we  have  to  take  into   account  the   processes
$l+\Phi^c\leftrightarrow   {\bar  l}  +  \Phi$   mediated   by  a
right-handed Majorana neutrino, as well as the processes:
\be
{\bar l}\leftrightarrow N\Phi^c.
\ee
The  Boltzmann  equation  for the  number  density  of the  light
left-handed leptons is:
\bea
{\dot n_l}+3 H n_l &=& \int\!\!{\rm d}\Pi_N\,{\rm d}\Pi_1
\,{\rm d}\Pi_2\,(2\pi)^4\,\delta^{(4)}(p_N-p_1-p_2)\nonumber\\
&&\times\left[-(1-\epsilon)f_l(p_1)f_{\Phi^c}(p_2)+(1+\epsilon)f_N(p_N)\right]|
{\cal M}_0|^2\nonumber \\
&&+2\int\!\!{\rm d}\Pi_1\,{\rm d}\Pi_2\,{\rm d}\Pi_3\,{\rm d}\Pi_4\,
(2\pi)^4\,
\delta^{(4)}(p_1+p_2-p_3-p_4)\nonumber \\
&&\times\left [ -f_l(p_1)f_{\Phi^c}(p_2)|{\cal M^\prime}
(l\Phi^c\to {\bar l}\Phi)|^2
\right.\nonumber \\
&&\left.+f_{\bar l}(p_3)f_{\Phi}(p_4)|{\cal{M^\prime}}({\bar l}\Phi\to l\Phi^c
)|^2
\right]
+{\cal O}(\epsilon n_l+n_l^2).
\eea
The origin of the various terms is  described below.

The first  interaction term comes from direct  $CP$-violation  in
the decay of the heavy Majorana  neutrinos and is proportional to
$\epsilon$,  the  $CP$-violating  phase  and the  squares  of the
matrix  element  $|{\cal  M}_0|$.  The second term  describes the
generation  of leptons from an initial  neutrino  state  $|N_i>$,
with number density $n_i$.  This converts into an antineutrino of
a different  generation $|N_j^c>$ which then decays into a lepton
and a higgs $\Phi$ with decay width:
\be
\Gamma_j={h^2_{\alpha j}\over 16\pi}.
\ee
The  last  term  takes  into  account  the  $2\leftrightarrow  2$
scattering mentioned above.  The corresponding equation for $\bar
l$ is obtained as usual, by interchanging $l\leftrightarrow {\bar
l}$, $\epsilon\leftrightarrow -\epsilon$, etc.
To  obtain  the   Boltzmann   equation   for  the   evolution  of
$n_L={1\over  2}(n_l-n_{\bar  l})$ we subtract  the  equation for
$n_{\bar  l}$ from that for  $n_l$ and  multiply  by a factor  of
$1/2$:
\be
{\dot
n}_L+3Hn_L=\epsilon\Gamma_i(n_i-n_i^{eq})-n_L(n_i^{eq}/n_\gamma)\Gamma_i
-2n_L n_\gamma <\sigma|v|>,
\ee
where $|{\cal  M}^\prime(l\Phi^c\to {\bar l}\Phi)|^2$ and $|{\cal
M}^\prime({\bar  l}\Phi\to  l\Phi^c)|^2$  are the  squares of the
matrix  elements  for   $2\leftrightarrow   2$  $L$-nonconserving
scatterings with the part due to real,  intermediate-state  $N$'s
removed.  Here the quantity:
\bea
<\sigma|v|>&=&\int\!{\rm d}\Pi_1{\rm d}\Pi_2{\rm d}\Pi_3{\rm
d}\Pi_4(2\pi)^4\delta^{(4)}(p_1+p_2-p_3-p_4)\nonumber\\
&&\qquad\qquad\qquad
f_l(p_1)f_l(p_2)|{\cal M}^\prime (l\Phi^c\to {\bar l}\Phi)|^2/n_\gamma^2,
\eea
is  the  velocity-averaged   $2\leftrightarrow  2$  $L$-violating
cross-section.  The  presence  of  the  term   $-\epsilon\Gamma_i
n_i^{eq}$  is  due  to  the   $CP$-violating   part  of  {$|{\cal
M}^\prime(l\Phi^c\to  {\bar  l}\Phi)|^2-|{\cal  M}^\prime  ({\bar
l}\Phi\to l\Phi^c)|^2$}.

In parallel to the calculation for the $n_i$, we obtain:
\be
{{\rm d}Y_L\over {\rm d}x}=\epsilon K\gamma x(Y_L-Y^{eq})-g_*Y^{eq}Y_L
K\gamma x-{{2Y_L\Gamma_s x}\over H(x=1)},
\ee
where  $\Gamma_s=n_\gamma<\sigma|v|>$, with the initial condition
$Y_L(0)=0$.

To solve the set of coupled  differential  equations, we take for
$\Gamma_i$, $\gamma$ and $\Gamma_s$:
\begin{equation}
\Gamma_i ={f^2\over {16\pi}}M_i\left\{
\begin{array}{ll}
x & x\ll 1 \\
1 & x\gg 1
\end{array}  \right. ,
\label{Gi}
\end{equation} 
\begin{equation}
\gamma =  \left\{
\begin{array}{ll}
x  & x \ll 1 \\
1 & x \gg 1
\end{array}  \right. , 
\label{gi}
\end{equation}
and:
\begin{equation}
\Gamma_s={3f^4\over {4\pi}} {M_i\over x}\left \{
\begin{array}{ll}
1 & x\ll 1 \\
{2\over 3}{1\over x^2} & x\gg 1
\end{array} \right. .
\label{Gs}
\end{equation}
The  functions  $\gamma$,  $\Gamma_i$,  and  $\Gamma_s$  must  be
specified in the region $x\approx 1$, in order to find a solution
for the differential equations.  We take:
\be
\gamma = 1-\exp (-x),
\ee 
\be
\Gamma_i={{M_i f^2}\over {16\pi}}\left ( 1-\exp (-x)\right ),
\ee
and:
\be
\Gamma_s={{3f^4 M_i}\over {4\pi x}}\left ( 1-\exp[-2/(3x^2)]\right ),
\ee
which  are a  good  approximation  to the  functions  (\ref{Gi}),
(\ref{gi})  and  (\ref{Gs})  above, in the  regions  $x\ll 1$ and
$x\gg 1$.  The numerical solutions to the Boltzmann equations are
then  obtained for different  values of $f$, $K$ and  $\epsilon$.
We summarize our observations below.

For $K \ll 1$, the amount of lepton  asymmetry grows cubically to
a constant  asymptotic value we call it $Y_L^{asym}(K  \ll 1)$ as
shown in figure  4.  In all  these  figures  we have  taken  some
representative  values for the couplings to be of the order of 1.
The  nature  of the curve is  independent  of the  choice  of the
Yukawa  couplings.  This asymptotic value is given by, $Y_L(asym)
=  {\epsilon  + \delta  \over  g_*}$.  This is the case  when the
decay  rate of  $N_1$  is less  than  the  expansion  rate of the
universe.  In this case the  scattering  rates are also less than
the  expansion  rate  of the  universe.  If the  mass  difference
between  $N_1$  and $N_2$ is of the same  order of  magnitude  to
their masses, the contributions of $\epsilon$ and $\delta$ become
comparable.  This  case has  been  discussed  extensively  in the
literature  and the  constraint on the scale of $(B-L)$  breaking
obtained from this  condition is $M_R > 10^7$ GeV with the Yukawa
couplings  to be of the  order  of  $10^{-5}$.  Somewhat  smaller
values of the Yukawa  couplings  can reduce  the scale of $(B-L)$
breaking  to a lower  value.  However,  then the amount of lepton
asymmetry  will  also  be  inadequate,   unless  there  is  large
hierarchy among the Yukawa couplings of different generations.

When the mass  difference  $|M_1 - M_2|  \sim  10^{-3}  M_{1,2}$,
$\delta$  can be three to four orders of  magnitude  larger  than
$\epsilon$.  In this case we can consider  $|f| \sim 10^{-7}$ and
still get  adequate  amount of lepton  asymmetry.  This will then
allow us somewhat  smaller right handed  symmetry  breaking scale
$M_R > 10^5$ GeV.

When $K$ is of the order of unity or more, the  lepton  asymmetry
vanishes  before $T = M_1$.  From $T = M_1$  onwards  the  lepton
asymmetry starts  increasing from its initial value of $Y_L = 0$,
but as it approaches the asymptotic  value of  $Y_L^{asym}(K  \ll
1)$  the  scattering   processes  becomes   important  and  start
depleting  it  exponentially  (as  shown in figures 5 and 6).
Unlike the common folklore that when the system is in equilibrium
the  asymmetry  falls  exponentially  fast,  soon the  scattering
processes become unimportant and the lepton asymmetry reaches its
new  asymptotic  value,  which is less than  ${\epsilon  + \delta
\over g_*}$.  For $K = 1$ the  suppression  factor is about $1/5$
and for  $K=5$  it is  .02.  Figure  7  shows  the  fall  of this
asymptotic values of the lepton asymmetry for different values of
$K$.  For  $K=1000$  the  suppression  factor  is about $8 \times
10^{-6}$.  In grand unified theories where the heavy gauge bosons
decay  generates  an $(B+L)$  asymmetry it was shown  \cite{kolb}
that this suppression  factor is $\approx [K ( {\rm ln} K )^{0.6}
]^{-1}$.  But in the case of  leptogenesis,  as  figure  7 shows,
this  suppression  is much  faster  than  linear.  For large mass
difference $|M_1 - M_2|$, $Y_L$ is approximately  proportional to
$f^2$ when all the $f$'s are of the similar order of magnitude and
$K \sim  10^{-3} f^2 {M_P \over  M_1}$.  Since $K$ and $Y_L$ both
are  proportional  to $|f|^2$, we cannot improve the bound on the
right handed  symmetry  breaking  scale with large $K$.  However,
since the suppression for large $K$ is almost quadratic we can at
most lower the scale of left-right symmetry breaking by one order
of magnitude to about $10^4$ GeV.

For very large $K$, if $|f|^4 \geq 10^{-17} M_1$, the  scattering
processes become larger than the expansion rate of the universe.
In this case the lepton  asymmetry  decreases  exponentially  and
never  reaches any  asymptotic  value.  In this region the actual
equilibrium condition is satisfied.

\section{Summary and Conclusion}

We  have  studied  the  possible  scale  of  left-right  symmetry
breaking,  which can  generate  enough  lepton  asymmetry  of the
universe.  For  simplicity  we assumed that the order of magnitude
of the Yukawa couplings of the heavy neutrinos are similar.
It was shown that the right handed  symmetry
breaking  scale  can be  lowered  to $10^4$  GeV at most.  Slight
deviation  from  the   out-of-equilibrium   condition  can  still
generate  similar  amount of leptons  asymmetry,  but that  cannot
change the scale of the symmetry breaking.

\section{Acknowledgment}
The work of S.L. is funded by a Marie Curie fellowship (TMR-ERBFMBICT-950565)
and that of J.F. and P.O'D by the Natural Sciences and Engineering Research Council of
Canada.

\newpage
\begin{figure}[htb]
\vskip 5in\relax\noindent\hskip -.5in\relax{\includegraphics{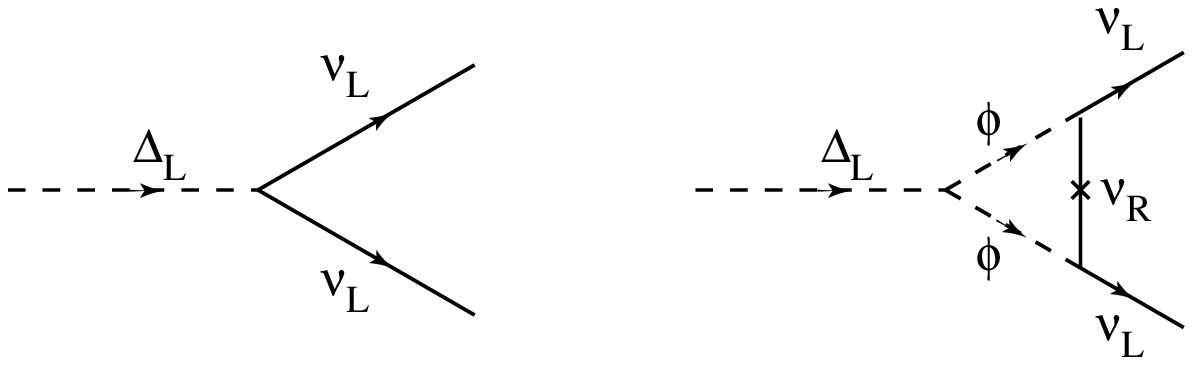}}
\caption{Tree and one loop diagrams of lepton number violating 
triplet higgs $\Delta_L$ decay}
\end{figure}
\newpage
\begin{figure}[htb]
\vskip 5in\relax\noindent\hskip -.5in\relax{\includegraphics{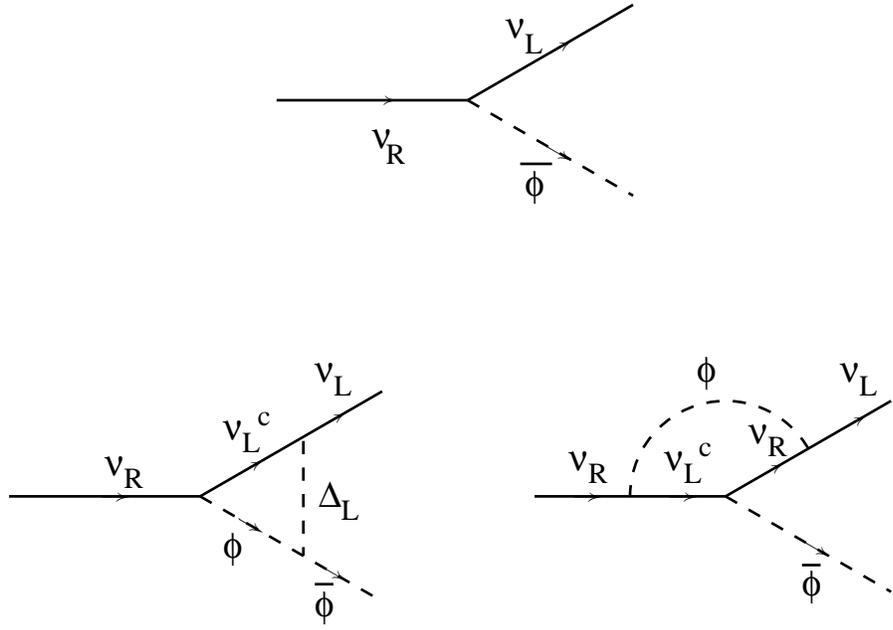}}
\caption{Tree and vertex correction type one loop 
diagrams contributing to the generation of lepton asymmetry}
\end{figure}
\newpage
\begin{figure}[htb]
\vskip 5in\relax\noindent\hskip -.5in\relax{\includegraphics{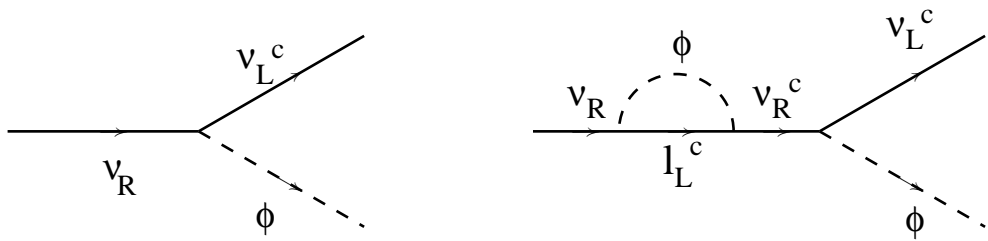}}
\caption{Tree and self energy correction type one loop 
diagrams contributing to the generation of lepton asymmetry}
\end{figure}
\begin{figure}[t]
\vskip 6.5in\relax\noindent \hskip -1.25in \relax{\includegraphics{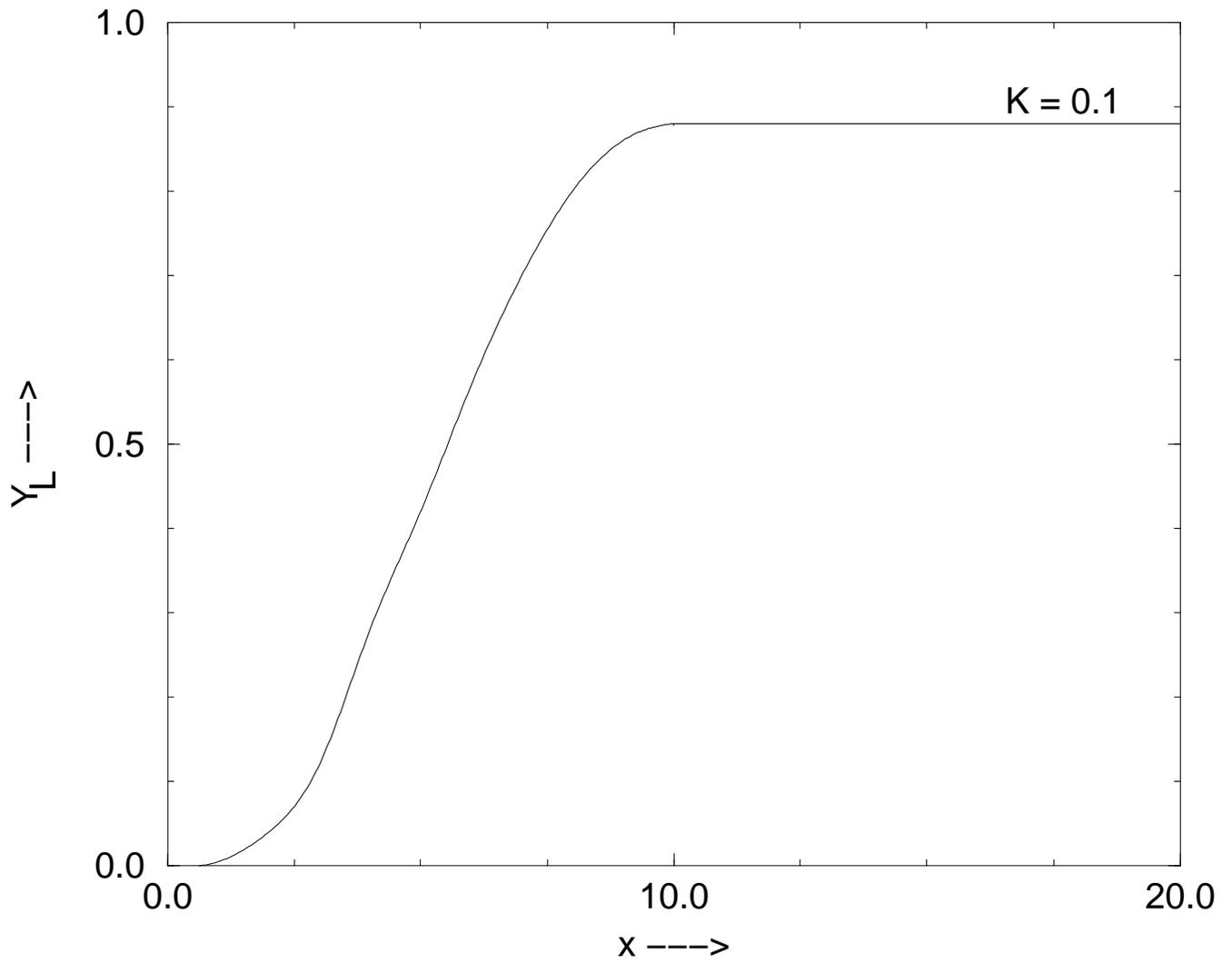}}
\caption{Lepton asymmetry $Y_L$ grows steadily to a constant asymptotic value
$\epsilon$ for $K < 1$. }
\end{figure}
\newpage
\begin{figure}[t]
\vskip 6.5in\relax\noindent\hskip -1.25in\relax{\includegraphics{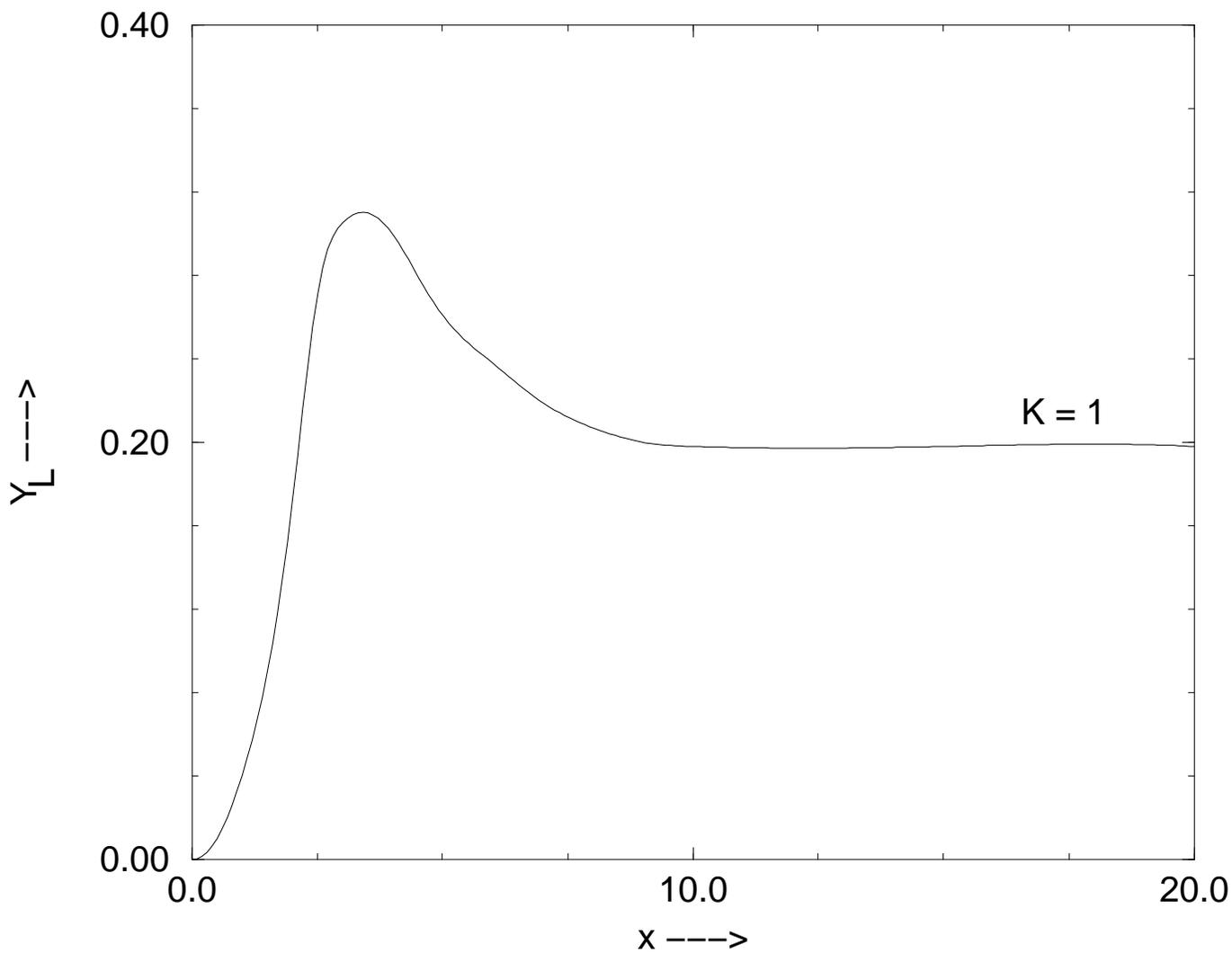}}
\caption{For $K = 1$, lepton asymmetry starts depleting before reaching a 
constant value. The asymptotic constant value is thus much less than $\epsilon$.
}
\end{figure}
\newpage
\begin{figure}[t]
\vskip 6.5in\relax\noindent\hskip -1.25in\relax{\includegraphics{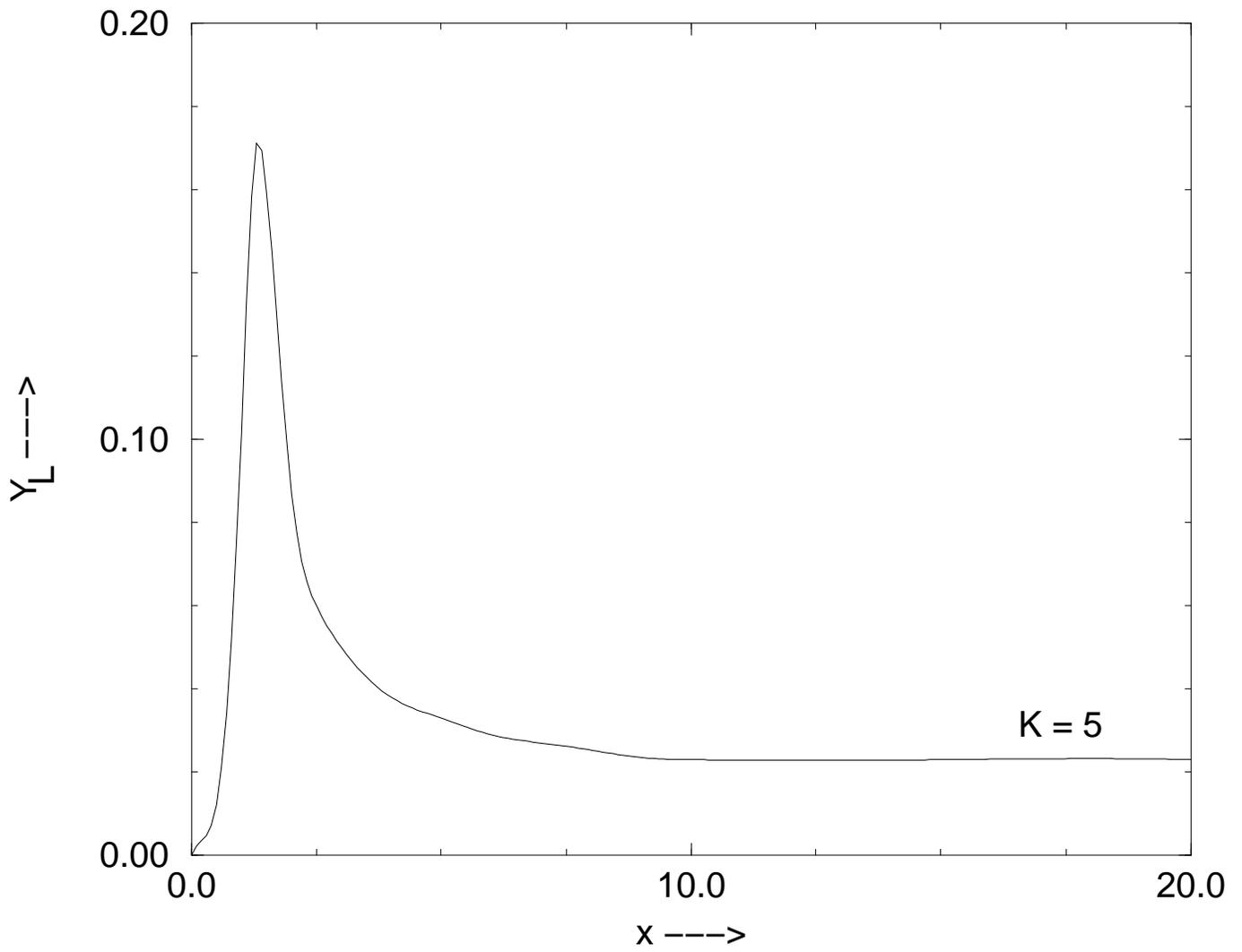}}
\caption{For large $K>1 $ the behaviour is similar to $K=1$. For $K = 5$
the asymptotic value is further depleted.
}
\end{figure}
\newpage
\begin{figure}[t]
\vskip 6.5in\relax\noindent\hskip -1.25in\relax{\includegraphics{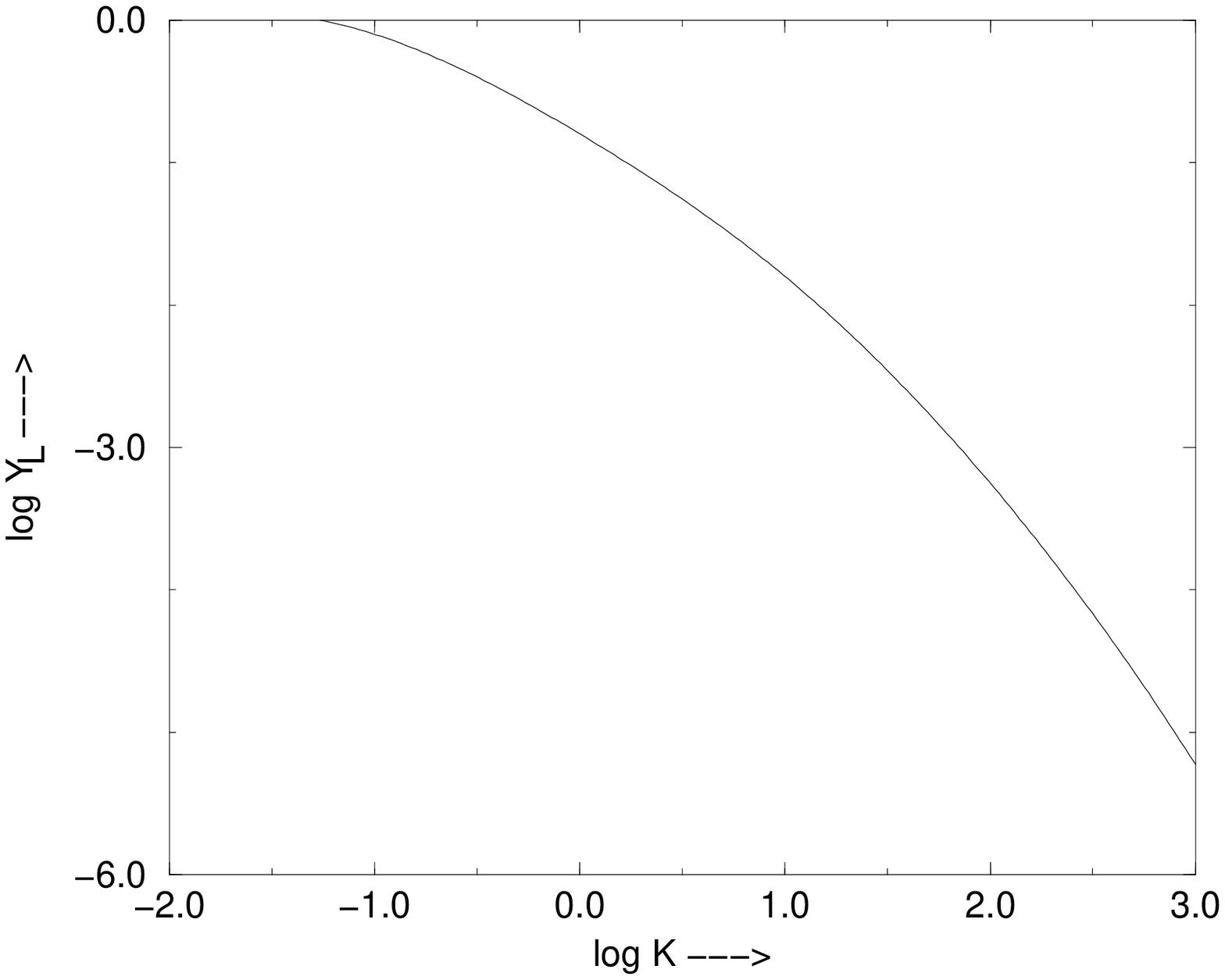}}
\caption{The asymptotic value of the lepton asymmetry for different values of 
$K$ in $log_{10} - log_{10}$ graph. For $K=1000$ the lepton asymmetry drops
to $8 \times 10^{-6}$.}

\end{figure}


\newpage
\begin{thebibliography}{99}  
%\baselineskip 18pt

\bibitem{sak} A.D. Sakharov, Pis'ma Zh. Eksp. Teor. Fiz. {\bf 5}
  (1967) 32.

\bibitem{kolb}E.W. Kolb and M.S. Turner, {\it The Early Universe} 
  (Addison-Wesley, Reading, MA, 1989).

\bibitem{gut} M. Yoshimura, Phys. Rev. Lett. {\bf 41} (1978)
  281; Erratum: Phys Rev. Lett. {\bf 42} (1979) 7461.

\bibitem{krs} V.A. Kuzmin, V.A Rubakov and M.E. Shaposhnikov,
  Phys. Lett. {\bf B 155} (1985) 36.

\bibitem{fy} M. Fukugita and T. Yanagida, Phys. Lett. {\bf B 174} 
  (1986) 45.

\bibitem{plum} M. Pl\"{u}macher, Z. Phys. {\bf C 74}, 549 (1997).

\bibitem{pat} P.J. O'Donnell and U.  Sarkar,  Phys.  Rev.  {\bf 
  D 49} (1994) 2118; W. Buchmuller and M. Pl\"{u}macher, Phys. 
  Lett. {\bf B 389}, 73 (1996).

\bibitem{luty} P.  Langacker, R.D.  Peccei and T.  Yanagida, Mod.
  Phys.  Lett.  {\bf  A  1}  (1986)  541;  M.A.  Luty, Phys.
  Rev.  {\bf D 45}  (1992)  455;  A.  Acker, H.  Kikuchi, E.
  Ma and U.  Sarkar,  Phys.  Rev.  {\bf D 48}  (1993)  5006.

\bibitem{mz} R.N.  Mohapatra  and X.  Zhang,
  Phys.  Rev.  {\bf D 46} (1992) 5331; A. Ganguly, J.C. Parikh 
  and U. Sarkar, Phys. Lett. {\bf B 385} (1996) 175.

\bibitem{ls}  A. Yu. Ignatev, V.A. Kuzmin and M.E. Shaposhnikov, JETP
  Lett. {\bf 30} (1979) 688; F.J.  Botella  and  J.
  Roldan,  Phys.  Rev.  {\bf D 44} (1991)  966; 
  J. Liu and G. Segre, Phys. Rev. {\bf D 48} (1993) 4609;
  L. Covi, E. Roulet and F. Vissani, Phys. Lett. {\bf B 384}, 169 (1996).

\bibitem{paschos} M. Flanz, E.A. Paschos and U. Sarkar, Phys. Lett.
  {\bf B 345} (1995) 248; M. Flanz, E.A. Paschos, U.  Sarkar 
  and J. Weiss, Phys. Lett. {\bf B 389}, 693 (1996).
  


\end{thebibliography}
\end{document}